# An integrated localization-navigation scheme for distance-based docking of UAVs

Thien-Minh Nguyen, Zhirong Qiu*, Muqing Cao, Thien Hoang Nguyen, and Lihua Xie, *Fellow, IEEE*

*Abstract*— In this paper we study the distance-based docking problem of unmanned aerial vehicles (UAVs) by using a single landmark placed at an arbitrarily unknown position. To solve the problem, we propose an integrated estimation-control scheme to simultaneously achieve the relative localization and navigation tasks for discrete-time integrators under bounded velocity: a nonlinear adaptive estimation scheme to estimate the relative position to the landmark, and a delicate control scheme to ensure both the convergence of the estimation and the asymptotic docking at the given landmark. A rigorous proof of convergence is provided by invoking the discrete-time LaSalle's invariance principle, and we also validate our theoretical findings on quadcopters equipped with ultra-wideband ranging sensors and optical flow sensors in a GPS-less environment.

## I. Introduction

Recent decade has witnessed a dramatic surge of small UAVs, with their wide applications in both civil and military areas, e.g. aerial photography, logistical delivery, surveillance, and disaster relief [1]. Essentially, the successful maneuver of a small UAV consists of solving two fundamental problems: localization and navigation, or estimation and control in a more general sense. These two problems are usually addressed in a separate manner. Most commonly is that an external localization system is assumed available to provide a reliable and accurate position estimate, e.g. global positioning system (GPS), or other positioning systems that require extra infrastructure [2], [3]. As a consequence, additional deployment and maintenance costs are also required for such systems, including a labor-intensive calibration, and it is highly demanding to establish such positioning systems in extreme environments. Moreover, these systems may suffer from a low flexibility with respect to environmental changes, as GPS is unavailable in city canyons. To ameliorate these issues, it is necessary to integrate localization and navigation into a combined framework.

As an initial effort to solve the integration problem, in this paper we study a distance-based docking problem for UAVs with a single landmark. More specifically, given a fixed landmark at an unknown and arbitrary location, we aim to navigate a UAV to the landmark by only using distance and odometry measurements. As an initial study, the trajectory planning of the UAV is generated from a discrete-time single integrator with bounded velocity. We propose an integrated localization-navigation scheme to solve the problem: a nonlinear adaptive estimation scheme to estimate the relative position to the landmark, and a delicate control scheme to ensure the convergence of the estimation, as well as the asymptotic docking at the given landmark. By employing techniques from adaptive estimation and the discrete-time LaSalles invariance principle, the efficacy of the overall localization-navigation scheme is rigorously established. Furthermore, experiments have also been conducted on quadcopters to validate the result, and it is promising to apply a similar scheme to a variety of practical scenarios.

Some related works can be found in [4]–[9], where the navigation of UAVs is based on distance measurements from landmarks at unknown positions. In summary, two kinds of navigation problems have been studied: a circumnavigation problem where a UAV is required to circle around a stationary or moving target [4]–[7], and a target pursuit or docking problem where a UAV is required to navigate to the prescribed position relative to the fixed landmark(s) [8], [9]. Different techniques have been proposed to solve these two problems. Specifically, for UAVs with unicycle dynamics, the distance measurements and the corresponding change rate were employed to tune the heading of the UAV to fulfill the navigation task [4]–[6]. On the other hand, based on adaptive estimation techniques [10], [11], certain kinds of trajectories were designed to simultaneously fulfill the localization and navigation tasks [7]–[9]. In comparison with the above works for continuous-time dynamics, our work considers a discrete-time formulation, which saves the trouble of gain tuning and is hence more convenient for implementation. Moreover, comparing with the relevant works [8], [9], we also consider the issue of input saturation which was not considered in [8], [9]. Finally, we implement the controller on quadcopters and conduct experiments to validate the theoretical findings.

The remainder of the paper is organized as follows: after stating the problem in Section II, we provide the integrated localization-navigation scheme in Section III. The corresponding convergence analysis is then conducted in Section IV. Simulation and experiment results are respectively provided in Section V and Section VI to validate the theoretical findings and demonstrate the practicality of the proposed algorithm. We conclude our work in Section VII.

Notations: in this paper we respectively use $\mathbf{N}$, $\mathbf{R}$ and $\mathbf{R}^+$ to denote the set of natural numbers, the set of real numbers, and the set of positive real numbers. For a vector $v \in \mathbf{R}^m$, $\|v\|$ and $\|v\|_\infty$ respectively stand for the Euclidean norm and the infinity norm, and $v'$ denotes its transpose.

The authors are with the School of Electrical and Electronic Engineering, Nanyang Technological University, 50 Nanyang Ave, 639798, Singapore.
*E-mail of coresponding author: qiuz0005@e.ntu.edu.sg

## II. PROBLEM FORMULATION

Given a fixed landmark at an arbitrary position $p^*$, we aim to dock a UAV to the landmark as follows:

$$\lim_{k\to\infty} p_k = p^*, \quad (1)$$

where $p_k \in \mathbf{R}^m$ denotes the position of the UAV at time step $k$. Specifically, for the purpose of trajectory planning, we consider a discrete-time integrator model with bounded velocity:

$$p_{k+1} = p_k + T\bar{u}_k, \|\bar{u}_k\|_\infty \le U, \quad (2)$$

where $T$ is the sampling period and $U$ is the maximum velocity. Clearly, for any control input $u_k$, the bounded velocity requirement can be satisfied by letting $\bar{u}_k = \pi_U(u_k)$, where $\pi_U(\cdot)$ is a projection operator onto $\bar{\mathcal{B}}(0,U)$ in $\mathbf{R}^m$ defined by

$$\pi_U(u_k) = \frac{U}{\max\{U, \|u_k\|\}} u_k \triangleq s_k u_k \quad (3)$$

## III. AN INTEGRATION SCHEME OF LOCALIZATION AND NAVIGATION

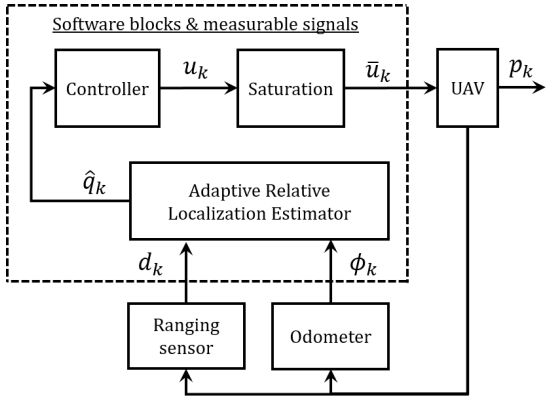

Fig. 1: An integrated localization-navigation scheme.

In this section we shall design an integrated localization and navigation scheme to simultaneously solve the relative localization and docking problem. To be detailed, as shown in Fig. 1, based on the distance measurement $d_k = \|p_k - p^*\|$, we shall construct an adaptive estimator to achieve the relative localization with the help of a specifically designed bounded input $\bar{u}_k$. Note that $\bar{u}_k$ should also be designed in a particular form to solve the docking problem (1). The design details of the estimator and the controller will be given in the next two subsections.

### A. Adaptive estimator for relative localization

Denote the relative position to the landmark by $q_k = p_k - p^*$. It is clear that by (2) we have

$$q_{k+1} = q_k + \phi_k, \quad (4)$$

where $\phi_k = T\bar{u}_k$ is the odometry measurement at time step $k$. Direct computation shows that

$$d_k^2 - d_{k-1}^2 = \|q_{k-1} + \phi_{k-1}\|^2 - \|q_{k-1}\|^2$$
$$= \|\phi_{k-1}\|^2 + 2\phi'_{k-1}q_{k-1}.$$

Now we can define the following parametric model

$$\zeta_k = \frac{1}{2}\left(d_k^2 - d_{k-1}^2 - \|\phi_{k-1}\|^2\right) = \phi'_{k-1}q_{k-1}. \quad (5)$$

Accordingly, the estimator for $q_k$ is constructed as follows:

$$\hat{q}_k = \hat{q}_{k-1} + \phi_{k-1} + \Gamma\phi_{k-1}\epsilon_k, \quad (6)$$

where $\epsilon_k = \zeta_k - \phi'_{k-1}\hat{q}_{k-1}$ and $0 < \Gamma < \alpha I$.

**Remark III.1.** *In [12] and [13] a similar parametric model was used to construct a continuous-time adaptive estimator for the relative localization. However, in the continuous-time model, the change rate of distance measurements is needed, and an additional time-varying observer is required, which is not the case for the discrete-time model in our work. Furthermore, both of the above works only considered the cooperative relative localization for multi-agents, and the navigation problem was not discussed.*

### B. Bounded controller for navigation

To solve the docking problem (1), we propose the following bounded controller:

$$\bar{u}_k = \pi_U(u_k), \ u_k = -\beta\hat{q}_k + Cf(d_k)\sigma_k, \ \beta > 0, \quad (7)$$

where $f$ and $\sigma_k$ are elaborated as follows.

$f : \mathbf{R}^+ \to \mathbf{R}^+$ is a function satisfying the following assumption:

**Assumption III.1.** $f(0) = 0$, and $0 < f(d) \le d, \forall d > 0$.

$\sigma_k \in \mathbf{R}^m$ is an external signal generated by an autonomous system as

$$\begin{aligned}\rho_{k+1} &= \Pi(\rho_k),\\ \sigma_k &= \Sigma(\rho_k), k = 0,1,2,\dots,\end{aligned} \quad (8)$$

where $\Pi$ and $\Sigma$ are two continuous mappings. Specifically, we require that $\sigma_k$ be constructed to satisfy the following assumption:

**Assumption III.2.**
1) *For any $\rho_0$, $\{\rho_k\}_{k=0}^\infty$ is bounded, and $\bar{\sigma} = \sup\{\|\sigma_k\|, k=0,1,\dots\} \le 1$.*
2) *There exists a constant $K$ such that $\sigma_{k+K} = -\sigma_k, \forall k \in \mathbf{N}$, and $\mathrm{span}\{\sigma_k : k=0,1,\dots,K-1\} = \mathbf{R}^m$.*

**Remark III.2.** *For the relative localization in 2D plane ($m=2$), we can select $\rho_0 \in \mathbf{R}^2$ to be a unit vector, $\Pi$ and $\Sigma$ to be matrix operators given by $\Sigma = I$ and $\Pi = \begin{bmatrix}\cos\omega & -\sin\omega\\ \sin\omega & \cos\omega\end{bmatrix}$ with $\omega = 2\pi/N$. Then it can be checked that Assumption III.2 is satisfied if $N \ge 4$ being an even integer. The general case of $m \ge 3$ will be discussed in the journal version of this work.*

**Remark III.3.** *Note that the controller design takes a similar form to [14], which consists of two terms: the first term is essentially a linear movement towards the landmark $p^*$ if there is no estimation error ($q_k = \hat{q}_k$), and the second one represents a circular movement (in 2D case) if $d_k \equiv d^* > 0$ with $\sigma_k$ generated by Remark III.2. Furthermore, in the absence of the first term ($\beta = 0$), the second term is also related*

*with the persistent excitation condition [15], and a large $C$ implies a faster convergence for the estimator. Based on the above observation, we may expect that $\Gamma$ and $\beta$ respectively influence the convergence of the control objective and the estimation, while $C$ is a tricky and important parameter for the performance of both the controller and the estimator. Section V will discuss more on the selection of parameters.*

### C. Main algorithm

We summarize the integrated localization-navigation scheme in the following algorithm:

---

**Algorithm 1** An integrated localization-navigation scheme

---
1: **Parameters**: $T, U, \Gamma, \beta, C, f(d), \Pi, \Sigma, \rho_0, d_\varepsilon$ (terminal distance)
2: **Initialization**: $\sigma := \Sigma\rho_0$, $\hat{q} := \hat{q}_0$, distance measurement $d_{now}$
3: **while** $d_{now} \geq d_\varepsilon$ **do**
    $u := -\beta\hat{q} + Cf(d_{now})\sigma$;
    **Generate control** $\bar{u} := \pi_U(u)$;
    $d_{old} := d_{now}$;
    Move the UVA by $T\bar{u}$ from the current position;
    Measure the displacement $\phi$ of the last movement;
    Measure distance $d_{now}$;
    $\zeta := \frac{1}{2}(d_{now}^2 - d_{old}^2 - \|\phi\|^2)$;
    $\epsilon := \zeta - \phi'\hat{q}$;
    **Update relative position estimate** $\hat{q} := \hat{q} + \phi + \Gamma\phi\epsilon$;
    $\rho := \Pi\rho$; $\sigma := \Sigma\rho$;
4: **end while**
5: Hover over or land on the landmark.

---

## IV. CONVERGENCE ANALYSIS

In this section, we shall provide a convergence analysis to show that the proposed localization-navigation scheme (6) and (7) can solve the docking problem (1). Firstly we shall show the stability, or boundedness, of the system states, then we show the convergence by invoking the LaSalle's invariance principle for discrete-time autonomous systems.

### A. Stability

In this section, we shall show the boundedness of estimation error $\tilde{q}_k = \hat{q}_k - q_k$ for relative localization, as well as that of the position error $q_k = p_k - p^*$ for navigation, respectively in Proposition IV.1 and Proposition IV.2. Specifically, we consider the following conditions:

$$\alpha(TU)^2 < 2 \quad (9)$$
$$C < \beta < 1/T, \quad (10)$$

where $\alpha, \beta, C$ are constants. Note that for fixed $\alpha$ and $C$, both of the above conditions can be met by a small sampling period $T$. On the other hand, for a fixed $T$, they can be satisfied by small $\alpha$ and $C$.

**Proposition IV.1** (Boundedness of estimation error). *Under the estimator (6) and condition (9), it holds that $\tilde{q}_k \in \ell_\infty$ and $\epsilon_k \in \ell_\infty \cap \ell_2$.*

*Proof.* Comparing (4) with (6), it is readily seen that

$$\tilde{q}_k = \tilde{q}_{k-1} - \Gamma\phi_{k-1}\epsilon_k = (I - \Gamma\phi_{k-1}\phi'_{k-1})\tilde{q}_{k-1}, \quad (11)$$

where we used the fact $\epsilon_k = -\phi'_{k-1}\tilde{q}_{k-1}$.

Define $V_k = \tilde{q}'_k\Gamma^{-1}\tilde{q}_k$ and $\Delta V_k = V_k - V_{k-1}$. Direct computation shows that

$$\Delta V_k = -\epsilon_k^2(2 - \phi'_{k-1}\Gamma\phi_{k-1}) \leq 0, \quad (12)$$

if we note that $(2-\phi'_{k-1}\Gamma\phi_{k-1}) > 0$ as a result of $\|\phi_{k-1}\| \leq TU$ and $\alpha(TU)^2 < 2$. Therefore, $V_k \leq V_0$, implying the boundedness of $\tilde{q}_k$. Moreover, since $\phi_{k-1}$ is bounded, it also implies that $\epsilon_k = -\phi'_{k-1}\tilde{q}_{k-1}$ is bounded. In summary, we have shown that $\tilde{q}_k, \epsilon_k \in \ell_\infty$.

To show that $\epsilon_k \in \ell_2$, denote $\lim_{k\to\infty} V_k = V_\infty \leq V_0$, and note that

$$0 \geq V_\infty - V_0 = \sum_{k=1}^\infty \Delta V_k = -\sum_{k=1}^\infty \epsilon_k^2(2 - \phi'_{k-1}\Gamma\phi_{k-1}).$$

The conclusion is readily achieved by noticing that $2 - \phi'_{k-1}\Gamma\phi_{k-1} > 2 - \alpha(TU)^2 > 0$. ∎

**Proposition IV.2** (Ultimate Boundedness of position error). *Under the bounded controller (7) and conditions (9) and (10), there exists a constant $M$ such that $\limsup_{k\to\infty} \|q_k\| \leq M$.*

*Proof.* Define $D_k = d_k^2$ and $\Delta D_k = D_k - D_{k-1}$. Recalling the definition of $\pi_U$ in (3), we obtain by (4) that

$$\begin{aligned}\Delta D_k &= (q_k + T\bar{u}_k)'(q_k + T\bar{u}_k) - q'_k q_k \\ &= s_k T(s_k T u'_k u_k + 2u'_k q_k),\end{aligned} \quad (13)$$

which gives rise to $\Delta D_k / s_k T \leq Tu'_k u_k + 2u'_k q_k \triangleq \Delta \tilde{D}_k$ as a result of $s_k \in (0,1]$. By substituting $u_k = -\beta(q_k + \tilde{q}_k) + Cf_k\sigma_k$ into $\Delta\tilde{D}_k$ with $f_k = f(d_k)$, we get that (we shall omit the subscript $(\cdot)_k$ to make the notation compact)

$$\begin{aligned}\Delta\tilde{D} &= T\beta^2(\|q\|^2 + \|\tilde{q}\|^2 + 2\tilde{q}'q) + TC^2 f_k^2 \|\sigma\|^2 \\ &\quad - 2T\beta Cf\sigma'(q+\tilde{q}) - 2\beta(\|q\|^2 + \tilde{q}'q) + 2Cf\sigma'q \\ &= (T\beta^2 - 2\beta)\|q\|^2 + TC^2 f_k^2 \|\sigma\|^2 \\ &\quad + (2 - 2T\beta)Cf\sigma'q + (2T\beta^2 - 2\beta)\tilde{q}'q \\ &\quad - 2T\beta Cf\sigma'\tilde{q} + T\beta^2\|\tilde{q}\|^2. \\ &\leq [T\beta^2 + TC - 2\beta + (2-2T\beta)C]\|q\|^2 \\ &\quad + 2\beta[(1-T\beta) + TC]\|q\|\|\tilde{q}\| + T\beta^2\|\tilde{q}\|^2, \quad (14)\end{aligned}$$

where to attain the last inequality we used the facts that $f_k \leq \|q_k\|$ in Assumption III.1, $\|\sigma_k\| \leq 1$ in Assumption III.2, $1 - T\beta > 0$, as well as the Cauchy-Schwartz inequality. By recalling (10), the coefficient of $\|q_k\|^2$ is given by

$$\begin{aligned}T\beta^2 + TC - 2\beta + (2-2T\beta)C &= (\beta - C)[T(\beta - C) - 2] \\ &\leq -(\beta - C)(TC + 1) < 0.\end{aligned}$$

Therefore, in combination with the boundedness of $\|\tilde{q}_k\|$ inferred from Proposition IV.1, we can find a constant $M$ such that $\Delta\tilde{D}_k < 0$ when $\|q_k\| \geq M$, or equivalently $\Delta D_k < 0$, which is the conclusion. ∎

## B. Convergence

With the preparation in the last section, we are ready to assert the convergence of $q_k$ to 0 by invoking the LaSalle's invariance principle for discrete-time autonomous systems (see Theorem 1 in [16]).

**Theorem IV.1.** *Under Assumptions III.1 and III.2, the distance-based docking problem (1) can be solved by combining the adaptive estimator (6) and the bounded controller (2) and (7), if we select proper gains to satisfy conditions (9) and (10).*

*Proof.* The overall system is given by combining the update protocol of $q_k, \tilde{q}_k, \rho_k$ respectively in (4), (11) and (8) as follows:

$$\begin{aligned} q_{k+1} &= q_k + \phi_k, \\ \tilde{q}_{k+1} &= (I - \alpha \phi_k \phi_k')\tilde{q}_k, \\ \rho_{k+1} &= \Pi(\rho_k), k = 0, 1, 2, \ldots, \end{aligned} \quad (15)$$

where $\phi_k = T\bar{u}_k = Ts_k u_k$, $u_k = -\beta(q_k + \tilde{q}_k) + Cf(\|q_k\|)\sigma_k$, and $\sigma_k = \Sigma \rho_k$. We have shown the boundedness of the system state by recalling Propositions IV.1 and IV.2, as well as 1) of Assumption III.2. By LaSalle's invariance principle, all trajectories will converge to the maximum invariant set $\mathcal{I}$ included in $\Delta V_k = 0, k \in \mathbf{N}^+$, where $V$ is the Lyapunov function defined in the proof of Proposition IV.1. We shall show that for any trajectory $\{[q_k', \tilde{q}_k', \rho_k']' : k = 0, 1, 2, \ldots\}$ in $\mathcal{I}$, it must hold that $q_k \equiv 0$.

Actually, it is readily seen from (12) that $\Delta V_k \equiv 0$ iff $\epsilon_k = -\phi_{k-1}'\tilde{q}_{k-1} \equiv 0$, which also follows that $\tilde{q}_k \equiv \tilde{q}_0$ by (11). Below we consider $\phi_k'\tilde{q}_0 \equiv 0$ for 3 cases to establish that $\phi_k'\tilde{q}_0 \equiv 0$ dictates either $\phi_k \equiv 0$ or $\tilde{q}_0 = 0$.

**Case 1**: $\tilde{q}_0 \neq 0$ and $\phi_k \not\equiv 0$. In this case, since

$$0 \equiv \phi_k'\tilde{q}_0 = (q_{k+1} - q_k)'\tilde{q}_0 = (\hat{q}_{k+1} - \hat{q}_k)'\tilde{q}_0,$$

we get that $\hat{q}_k'\tilde{q}_0 \equiv \hat{q}_0'\tilde{q}_0$. On the other hand, $0 \equiv \phi_k'\tilde{q}_0$ also implies that $0 \equiv u_k'\tilde{q}_0 = [-\beta\hat{q}_k + Cf(d_k)\sigma_k]'\tilde{q}_0$, or equivalently $\beta\hat{q}_0'\tilde{q}_0 \equiv \beta\hat{q}_k'\tilde{q}_0 \equiv Cf(d_k)\sigma_k'\tilde{q}_0$. Specifically, we have $f(d_k)\sigma_k'\tilde{q}_0 = f(d_{k+K})\sigma_{k+K}'\tilde{q}_0$. If $q_k = 0$, then $d_k = 0$ and $u_k = -\beta\tilde{q}_k = -\beta\tilde{q}_0$, and $\phi_k = Ts_k u_k = -\beta Ts_k\tilde{q}_0$, which follows that $\phi_k'\tilde{q}_0 = -\beta Ts_k\|\tilde{q}_0\|^2 \neq 0$, a contradiction. Therefore, $f(d_k) > 0, \forall k \in \mathbf{N}$.

On the other hand, if we remember $\sigma_{k+K} = -\sigma_k$ in Assumption III.2, we can further obtain that

$$[f(d_k) + f(d_{k+K})]\sigma_k'\tilde{q}_0 \equiv 0, k = 0, 1, \ldots, K-1. \quad (16)$$

As a consequence of $f(d_k) > 0$ for any $k$, the above can be simplified as $\sigma_k'\tilde{q}_0 \equiv 0$. In addition, noticing that span$\{\sigma_k : i = 0, 1, \ldots, K-1\} = \mathbf{R}^m$ in Assumption III.2, we can find a linear combination of $\tilde{q}_0$ as $\tilde{q}_0 = \sum_{k=0}^{K-1} a_k \sigma_k$, which follows by (16) that $\|\tilde{q}_0\|^2 = \tilde{q}_0' \sum_{k=0}^{K-1} a_k \sigma_k = 0$, another contradiction.

In summary, $\phi_k'\tilde{q}_0 \equiv 0$ dictates that $\phi_k \equiv 0$ or $\tilde{q}_0 = 0$.

**Case 2**: $\phi_k \equiv 0$. In this case, $u_k \equiv 0$ and $q_k \equiv q_0$, which yields that $d_k \equiv d_0$ and $\hat{q}_k \equiv \hat{q}_0$. Since $u_k = -\beta\hat{q}_k + Cf(d_k)\sigma_k \equiv 0$, we have $Cf(d_0)\sigma_k \equiv \beta\hat{q}_0$. By remembering that span$\{\sigma_k : k = 0, 1, \ldots, K-1\} = \mathbf{R}^m$ in Assumption III.2, the only possible case is that $d_0 = 0$, namely $q_k \equiv 0$.

**Case 3**: $\tilde{q}_0 = 0$. In this case, the estimation error is always 0, and the relative position dynamics (4) is simplified as

$$q_{k+1} = q_k + Ts_k(-\beta\hat{q}_k + Cf(d_k)\sigma_k), \quad (17)$$

where $s_k$ was defined in (13). It is clear that $q^* = 0$ is a globally asymptotically stable equilibrium for (17), if we note that $\Delta \tilde{D}_k \leq [T\beta^2 + TC - 2\beta + (2 - 2T\beta)C]\|q_k\|^2$ in (14).

In summary, we have completed the proof. ∎

## V. SIMULATION

In this section, we will verify the feasibility of the localization-navigation Algorithm 1 by numerical simulation under different settings. Firstly for a static landmark, we will examine the performance under different gains $\alpha, \beta$ and $C$, and select a proper set of gains to achieve good performance in the experiment in Section VI. Furthermore, we will also show the simulation result for a slowly drifting landmark. To be consistent with the experiment setup, we are only concerned with the 2D case, and select the sampling period $T = 0.1s$, maximum velocity $U = 0.75m/s$, and $f(d) = d$. Moreover, we select $\Gamma = \alpha I$ as the estimator gain matrix (with some notation abuse), and generate $\sigma_k$ by Remark III.2 with $\rho_0 = [1, 0]'$ and $N = 36$. For each simulation we always assume that the initial estimate of the relative position is given by $\hat{q}_0 = [0, 0]'$, and the UAV starts from the relative position $q_0 = [-25, -32]'$.

*1) A static landmark:* In this section we are concerned with a static landmark. Under the above setting, we can select positive constants $\alpha, \beta$ and $C$ as follows to solve the docking problem by invoking conditions (9) and (10) in Theorem IV.1:

$$\alpha < 355.56, C < \beta < 10. \quad (18)$$

To examine the performance under these different gains, we have done different simulations by only changing one gain each time, as shown from top to down in Figure 2. Note that the red, blue and green lines respectively show the relative position $q_k$, estimated relative position $\hat{q}_k$, and the estimation error $\tilde{q}_k$.

Clearly, the docking problem is solved in all 4 cases as condition (18) is satisfied. Comparing Fig. 2a and Fig. 2b, we can see that a larger estimator gain $\alpha$ leads to a faster localization convergence. If we further increase the controller gain $\beta$ to 5, then we observe a faster convergence to the landmark, but with a slower estimator convergence and a more oscillating trajectory, as shown in Fig. 2d. However, if we increase the excitation magnitude $C$ from 1 to 4, then not only the fast localization can be recovered, but the docking problem can also be solved fastly with a smooth trajectory, as shown in Fig. 2d which achieves the best performance among 4 cases in terms of fast localization and navigation. Note that the above process also shed some light on a proper gain tuning for better performance of the integrated scheme.

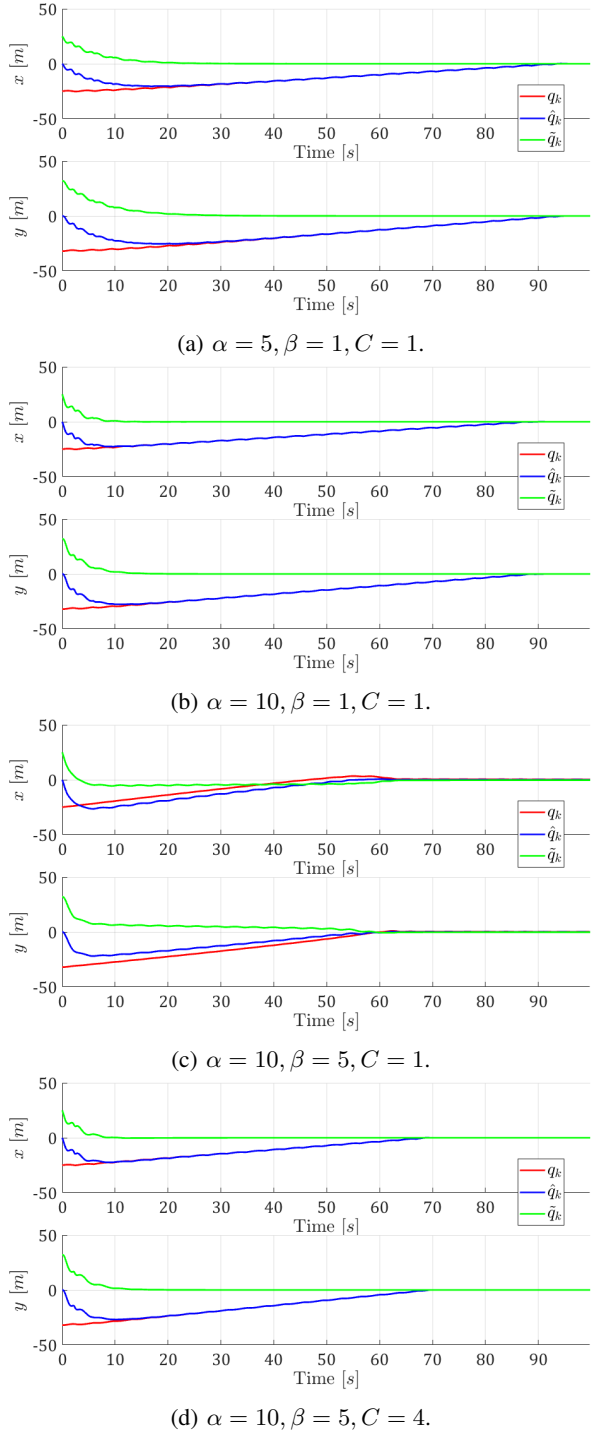

(a) $\alpha = 5, \beta = 1, C = 1$.

(b) $\alpha = 10, \beta = 1, C = 1$.

(c) $\alpha = 10, \beta = 5, C = 1$.

(d) $\alpha = 10, \beta = 5, C = 4$.

Fig. 2: Performance comparison under different gains.

*2) A slowly drifting landmark:* To further demonstrate the capability of the algorithm, in this part we consider a slowly drifting landmark and apply the Algorithm 1 with $\alpha = 10$, $\beta = 5$, $C = 4$ as selected in the last section. Specifically, we assume that the dynamics of the landmark is given by

$$p_k^* = p_0^* + \begin{bmatrix} 5\cos(kw) \\ 5\sin(kw) \end{bmatrix} + kT \begin{bmatrix} -0.01875 \\ 0.00375 \end{bmatrix}, \quad (19)$$

where $w = 1.534 \times 10^{-3}$ and $T = 0.1$. We can see from Fig. 3 that the agent is still able to track this moving target.

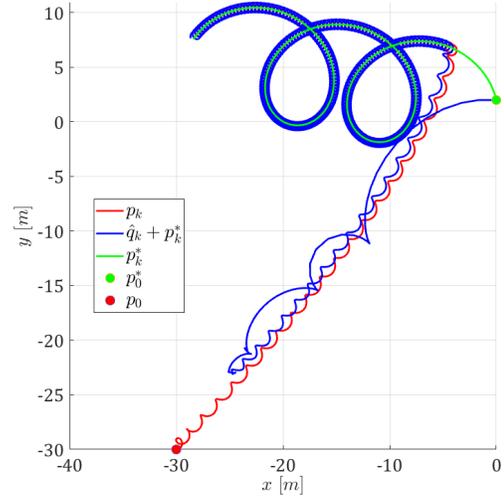

Fig. 3: Simulation for a slowly drifting target. The target's initial location is marked by the green circle and the UAV's intial location is marked by the red circle.

## VI. EXPERIMENTS ON QUADCOPTERS

To further validate our theoretical findings in real flights, we have implemented the integrated localization-navigation Algorithm 1 on quadcopters and conducted multiple tests in a GPS-less environment. To be detailed, the algorithm is executed in real-time on an on-board computer running Ubuntu and Robot Operating System (ROS). Noticing that UWB is robust to multipath and non-line-of-sight effects, and provides a reliable long distance ranging with an error within only a few centimeters [3], [17], [18], we obtain distance measurements by using two UWB nodes, with one mounted on a so-called *target UAV* hovering at the destination, and the other one installed on the *autonomous UAV*.

To obtain the odometry measurements, we fuse the output from an optical flow sensor [19] with the measurements from an on-board altimeter and inertial measurement unit. The values of the parameters of the Algorithm 1 are exactly the same to the ones used in the simulation of Fig. 2d.

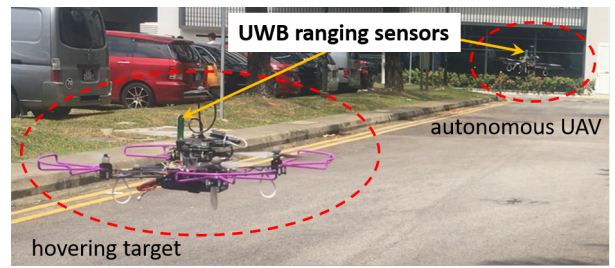

Fig. 4: Experiment setup.

The tests were conducted in a $20m \times 30m$ large open area without using GPS. As shown in Fig. 4, the landmark is fixed on a UAV stably hovering at some unknown position, and the autonomous quadcopter aims to approach the landmark from a distant place by using distance measurements. Since we are concerned with the experiment of 2D case for validation purpose, we use the planar distance $d_k = \sqrt{r_k^2 - q_{k,z}^2}$, where $r_k$ is the UWB ranging measurement, and $q_{k,z} = p_{k,z} - p_z^*$ is the relative height to the landmark ($p_z^*$ is transmitted from the target to the autonomous UAV via a zigbee communication network). A similar revision also applies to the odometry measurements $\phi_k$. Furthermore, to prevent the quadcopter from colliding with the landmark, we set the terminal distance $d_\varepsilon = 1.5$, namely that the quadcopter will stop to hover when the mutual distance is within 1.5 m. Video recording of one flight test can be viewed at https://youtu.be/2eyNzXXAhLM.

A total of 5 tests with different starting points have been conducted to demonstrate the capability of the integrated localization-navigation scheme. To evaluate the experiment results, we plot the planar distance decrease of all tests in Fig. 5, and it is readily seen that in all cases the planar distance decreases steadily and quickly until the terminal distance is met. In conclusion, the experiment results have shown a competitive performance of the integrated localization-navigation scheme, and it is promising to further apply a similar integration idea to more practical scenarios.

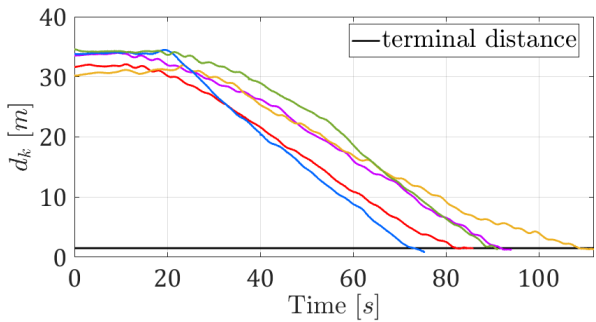

Fig. 5: A quick and steady docking to the landmark.

## VII. CONCLUSION AND FUTURE WORKS

In this paper we studied the distance-based docking problem of UAVs by using a single landmark placed at an arbitrarily unknown position. An integrated estimation-control scheme was proposed to simultaneously achieve the relative localization and navigation tasks for discrete-time integrators under bounded velocity: a nonlinear adaptive estimation scheme to estimate the relative position to the landmark, and a delicate control scheme to ensure both the convergence of the estimation and the asymptotic docking at the given landmark. By invoking the discrete-time LaSalle's invariance principle, we showed that the docking problem can be solved by selecting proper gains of the estimator and the controller. Simulation under different settings was conducted to show the performance of the algorithm, and the theoretical findings were also validated in a 2D GPS-less environment by implementing the integrated scheme on quadcopters equipped with ultra-wideband ranging sensors and optical flow sensors. More general cases for UAVs with other kinds of system dynamics are under investigation.

ACKNOWLEDGEMENT

We also appreciate the great help from Mr. Minh-Chung Hoang in building the quadcopter platform. We also want to thank Mr Han Wang for assisting us in the experiments.